# Can we Trust Chatbots for now?
# Accuracy, reproducibility, traceability; a Case Study on Leonardo da Vinci's Contribution to Astronomy

Didier EL BAZ

LAAS-CNRS, Université de Toulouse, CNRS, France

elbaz@laas.fr

**Abstract:** Large Language Models (LLM) are studied. Applications to chatbots and education are considered. A case study on Leonardo's contribution to astronomy is presented. Major problems with accuracy, reproducibility and traceability of answers are reported for ChatGPT, GPT-4, BLOOM and Google Bard. Possible reasons for problems are discussed and some solutions are proposed.

1. Introduction

Large Language Models (LLM) are a new area of research and potentially have many applications like chatbots, content creation, education... LLMs have received considerable attention. As an introduction to natural language processing, language models and n-grams, the reader is referred to the book by I. Goodfellow, Y. Bengio, and A. Courville [1]. Reference is also made to the recent book by L. Tunstall, L. von Werra and T. Wolf, which deals with natural language processing with transformers [2]. Several LLMs like OpenAI GPT-3, GPT-4 and BigScience BLOOM are currently used in chatbots, see [3], [4] and [5], respectively. Reference is also made to [6] to [9] for various applications of transformers.

Amongst LLMs, ChatGPT has recently reached a huge audience. It leverages on GPT-3.5, which is a variant of the GPT-3 model [3]. ChatGPT makes use of machine learning to treat natural language inputs and produce answers [10]. ChatGPT uses a combination of neural networks and deep learning techniques to analyze and process the patterns and structures in the language inputs. More precisely, the model is based on a transformer architecture that has been optimized for natural language processing. When a user submits a text to ChatGPT, the model first processes the input with the purpose to perceive its intended meaning. It then generates an answer based on the learned patterns and structures in the training data. ChatGPT rests on a pre-trained model that has been subsequently trained on a huge corpus of text data, a process called supervised learning. A peculiarity of ChatGPT is to have been trained partly using deep reinforcement learning with human feedback.

ChatGPT is multilingual. It can answer very fast to simple questions. In many cases, it can quickly provide very well-written chats with correct information for a wide range of fields including hot scientific topics. In Mathematics, ChatGPT has an excellent success rate when it comes to finding the next number in a series of numbers. Aspects related to ethics seem also to have been the subject of particular attention when training the large language model. I had the opportunity to test ChatGPT with discriminatory messages in a professional context for which I have obtained noteworthy results. I have also noted possible associations of concepts related to the same word; though this may be fortuitous. Several practical cases have been reported where ChatGPT can produce interesting studies in relation to legal issues or correct drafts in domains such as education and business. One novelty of GTP3 and ChatGPT is to be able to respond to confirmation requests. This feature is particularly important since despite impressive results that have been widely reported, it is questionable whether ChatGPT and other chatbots are reliable and can provide reproducible and traceable results. In this paper, we start addressing the above question. In particular, we present a small case study that deals with the contribution of Leonardo da Vinci to astronomy. The merit of this small case study is to highlight very important problems with the different chatbots we consider. Indeed, ChatGPT, GPT-4, BLOOM and Bard make colossal mistakes that call into question the entire history of science by ignoring the major contributions of Nicolaus Copernicus and Galileo Galilei, two of the greatest astronomers in history. We give elements in order to understand the origins of the mistakes and make suggestions to improve chatbots that would be specially designed for educational purpose.

Section 2 deals with the case study. Problems with accuracy, reproducibility and traceability of answers generated by ChatGPT and GPT-4 are reported and analyzed; possible reasons for problems are discussed. In section 3, we present and analyze answers on the same topic that were generated via BigScience BLOOM, Google Bard and FLAN-T5 XXL. Section 4 deals with a general analysis and presentation of possible solutions. Conclusions and future works are presented in Section 5.

2. Case study with OpenAI ChatGPT and GPT-4

Leonardo's contribution to astronomy is not an uninteresting or trivial topic. After more than five centuries, interest in Leonardo's work never waned. Leonardo's multi-faceted genius and talent has considerably attracted the attention of a large audience. Over time, many scientific or popular articles have been written; many serious studies and biographies have been released as well. As a consequence, there is a mass of data on Leonardo. I underline that much of the data was produced by Leonardo himself. Leonardo is certainly one of the Renaissance artists whose writings and drawings have been preserved the most. Unfortunately, many writings, drawings and even some masterpieces were also lost. Leonardo is well known for his contributions to anatomy and physiology, e.g., see [11] and [12], respectively. His knowledge of the subject had been acquired mainly by the dissection of corpses and collaboration with the anatomist Marcantonio della Torre in 1510. Leonardo's drawings and writings on that topic have gained considerable attention. In this domain, Leonardo was certainly ahead of his time by several decades. His contribution to hydrology is also widely recognized, e.g. see [13], [14] and [15].

On what concerns astronomy, the situation is somewhat different. Leonardo's manuscripts contain many pages, with notes and drawings on astronomy, e.g. see [16], [17] and [18]. Sometimes, his notes are contradictory. Scholars and experts such as Martin Kemp, Daniel Arasse, Vincent Delieuvin and Louis Frank convene that Leonardo was mostly interested in the light of celestial bodies and the way light is transmitted from a celestial body to another, see [19], [20] and [21]. We note that some popular papers have overestimated Leonardo's scientific contributions. Thus, I have decided to concentrate on that specific topic in order to evaluate the quality, reproducibility and traceability of answers. Of course, I do not intend to rank chatbots on a single question.

In the beginning of February 2023, my first question on that topic to ChatGPT was:

What are Leonardo da Vinci's contributions to astronomy?

It just took a couple of seconds for ChatGPT to write:

*"Although Leonardo da Vinci is primarily known for his achievements in art, engineering, and anatomy, he also made some contributions to astronomy. Here are a few examples:*

*Lunar Craters: Leonardo was one of the first people to observe and record the irregularities of the moon's surface, including the existence of craters. He was able to make these observations using a telescope, which had recently been invented.*

*Solar System Models: Leonardo was also interested in creating models of the solar system, and he created several different versions throughout his life. In some of these models, he proposed that the planets moved in circular orbits around the sun, rather than the more commonly accepted idea at the time of epicycles.*

*Geology and Fossils: In addition to his observations of the moon, Leonardo also made a number of observations about the Earth, including the formation of mountains and the presence of fossils. He recognized that the Earth had undergone significant changes over time and that the fossils he found were evidence of past life.*

*It's worth noting that many of Leonardo's ideas about astronomy were not widely accepted during his lifetime and were not influential in the development of modern astronomy. However, his curiosity and willingness to observe and experiment were important traits that are still valued by scientists today."*

I will skip the statement concerning geology and fossils since this point is somewhat off topic for a question that is dedicated to astronomy. Nevertheless, I note that the statement on geology and fossils, that belongs to the general domain of sciences of the universe, is certainly the most accurate. Unfortunately, the other statements are wrong. Today, highly reputed scholars convene that Leonardo was the first to give the explanation for the ashy light of the moon [21]. Leonardo explained in the Codex Leicester 2r, ca. 1506–1510, that the ashy light of the moon is due to the reverberation of the earth, i.e., the earthshine phenomenon [16]. I note that ChatGPT does not provide such an answer.

Telescopes are only considered on a conceptual point of view in Leonardo's drawings and writings. He writes: "*Make glasses to see the moon big*" in Manuscript C. A. 518 r, ca. 1513-1514 (see [22]). To the best of our knowledge, which are based on the recent research works on Leonardo's writings, nowhere did Leonardo write that he observed the moon and its craters via a telescope, nor did he write that he was able to make observations on the irregularities of the moon using a telescope. Similarly, we get no detailed drawing of the moon. In addition, glasses were not really operational until the early seventeenth-century. On what concerns significant improvements to the telescope and subsequent detailed observations of the moon (including drawings), Venus and satellites such as satellites of Jupiter, the credit goes unanimously to Galileo Galilei in the early seventeenth century [23].

Nowhere did Leonardo write a study on the trajectory of the planets around the sun. Scholars and experts in Leonardo studies do not point out any new contribution of Leonardo to celestial mechanics. He did not propose any new model on trajectory of celestial bodies. Leonardo was faithful to the Ptolemaic system that describes the trajectories of celestial bodies according to a geocentric model

(see [19]) as attested by several notes on the geocentric system like in the Codex Leicester, 2r ca. 1506–1510 (see [16]). Several entries in the list of his books in the second Codex of Madrid, 2v – 3r, ca. 1503 – 1505 refer to Ptolemy and the geocentric model (see [19] and [24]). Of course; there is the famous sentence: "*il sole non si muove*", i.e., the sun does not move, in the manuscript W 912669v of the Royal Library (see [13]). Nevertheless, this sentence is not supported by any demonstration. It proves nothing with regards to the real intention of Leonardo. We know that this statement is also a mistake since the sun orbits around the center of our galaxy.

The reader is referred mainly to the work of Aristarchus of Samos in the third-century BC for early studies on heliocentric systems in the ancient times. Nevertheless, heliocentrism did not receive a wide audience during Antiquity, Middle Age and early Renaissance. Geocentrism was commonly accepted at that time. In particular, the geocentric model of Ptolemy prevailed from the second-century AD till Renaissance. In the beginning of the sixteenth-century, Nicolaus Copernicus presented to his friends a hand-written treatise on a heliocentric model known as the "*Commentariolus*". The heliocentric model of Copernicus was finally published in 1543 in the book "*De revolutionibus orbium coelestium*", see [25]. Copernicus book is unanimously recognized as a major advance in astronomy and in the history of civilization.

Thus, ChatGPT makes very important mistakes which are in contradiction with history of science. In particular, the errors call into question the major contributions of two of the greatest astronomers in history: Nicolaus Copernicus and Galileo Galilei.

During the same session, when asked to confirm, ChatGPT persists in its errors and answers: "Yes, I can confirm my answer." When also asked on reference, ChatGPT answer is surprising. Below is the *verbatim* of my question and ChatGPT answer.

Question: What is you reference when you say: "Although Leonardo da Vinci is primarily known for his achievements in art, engineering, and anatomy, he also made some contributions to astronomy. Here are a few examples: Lunar Craters: Leonardo was one of the first people to observe and record the irregularities of the moon's surface, including the existence of craters. He was able to make these observations using a telescope, which had recently been invented. Solar System Models: Leonardo was also interested in creating models of the solar system, and he created several different versions throughout his life. In some of these models, he proposed that the planets moved in circular orbits around the sun, rather than the more commonly accepted idea at the time of epicycles?"

ChatGPTs: "*My response is based on the available historical records and scholarship on Leonardo da Vinci. The information about his contributions to astronomy, specifically his observations of the moon's surface and his proposed models of the solar system, is widely documented in various historical accounts, including biographies of Leonardo da Vinci and studies of his notebooks and drawings. Some examples of these sources include "Leonardo da Vinci and the Idea of a Solar System" by Edoardo Zanon and "Leonardo da Vinci and Astronomy" by Erwin Panofsky. These sources provide a more detailed and comprehensive overview of Leonardo da Vinci's contributions to astronomy.*"

Later, when questioned about the above quoted references ChatGPT writes: "*Leonardo da Vinci and Astronomy" by Erwin Panofsky was originally published as an article in the journal The Metropolitan Museum of Art Bulletin in 1938. The article was later included in the book Meaning in the Visual Arts by Panofsky, which was first published in 1955. The book has been reprinted many times since then and is widely available in both print and digital formats. It's worth noting that The Metropolitan Museum of Art Bulletin is a publication of the Metropolitan Museum of Art in New York City, which features articles about art and art history.*"

Unfortunately, I couldn't find any Erwin Panofsky's paper in the 1938 issues of The Metropolitan Museum of Art Bulletin, nor did I find information regarding a possible note of Leonardo on celestial mechanics, or more generally, any reference to a scientific contribution of Leonardo to astronomy, in the famous book of Erwin Panofsky: "*Meaning in the Visual Arts*", see [26]. Similarly, I couldn't find any paper or contribution about *"Leonardo da Vinci and the Idea of a Solar System"* by Edoardo Zanon that would deal with heliocentric model.

I highlight that ChatGPT's answers are not reproducible. The answer to the same question on the contributions of Leonardo da Vinci to astronomy varies according to time and sessions. Another day and in a different session, ChatGPT gives the following answer.

*"Leonardo da Vinci's contributions to astronomy include the following: 1. He was the first to propose that the Earth and other planets revolve around the Sun in 1514. 2. He developed the first scientific theories of the moon's phases, eclipses, and tides. 3. He was the first to use a telescope to study the stars, and made observations of the Milky Way and other astronomical phenomena. 4. He proposed a heliocentric model of the solar system and wrote on the importance of observing the stars. 5. He worked on a project to measure the distance between Earth and the Moon. 6. He developed a system of star charts and was the first to map the sky."*

In a different session, I got also the following answer.

*"Leonardo da Vinci, the famous artist, inventor, and polymath, made several contributions to the field of astronomy. Although he is not primarily known for his astronomical work, his notes and drawings contain many observations and speculations about the cosmos. Some of his most significant contributions include:*

*Moon observations: Da Vinci made detailed drawings of the Moon's surface, including craters and mountains. He also noted that the Moon has no atmosphere and suggested that the irregularities on its surface were caused by volcanic activity.*

*Solar System: Da Vinci speculated about the nature of the Solar System and the position of the Earth within it. He proposed that the planets revolve around the Sun and that the Earth is not the center of the universe, a revolutionary idea in his time.*

*Comet observations: Da Vinci made detailed observations of a comet in 1472, noting its brightness, tail, and changing position in the sky. He also speculated about the nature of comets and suggested that they might be made of gases.*

*Star observations: Da Vinci made many observations of stars, noting their brightness, color, and position in the sky. He also speculated about the nature of stars and their relationship to the Sun.*

*Astronomical instruments: Da Vinci designed several instruments for astronomical observation, including a telescope and a sextant.*

*Overall, while Leonardo da Vinci's contributions to astronomy may not be as well-known as his contributions to art and engineering, his observations and speculations helped to advance the understanding of the cosmos in his time and laid the groundwork for future astronomical discoveries."*

Parts of the last two statements like observation of stars (without telescope of course) and speculation about their nature are correct, e.g., see Manuscripts F 5 r and 4 v, ca 1508 – 1509, [27]; but the statements about heliocentrism and observations of the moon or the universe via a telescope are incorrect. The first and last answer also present contradictions. In the first answer Leonardo's ideas "*were not influential in the development of modern astronomy*". In the last chat "*his observations and*

*speculations helped to advance the understanding of the cosmos in his time and laid the groundwork for future astronomical discoveries."* No human evaluators will have difficulty distinguishing the above answers generated by ChatGPT from comments written by scholars since the answers are absurd or contain huge mistakes. I believe that the mistakes come from the fact that data used to train ChatGPT were not generated by scholars. They were probably collected on the web and correspond to possible contradictory data generated during a long period of time with questionable references. I believe also that there was insufficient training on this particular topic. It is highly probable that many popular topics which are associated to mass of data which was generated during a long period of time and which present sometimes contradictions give raise to the problem pointed in this paper.

On April 5, 2023, I tested GPT-4 on the same topic. GPT-4 is at the time of writing this article, the latest LLM developed by OpenAI [4]. The answer was slightly better: GPT-4 no longer refers to observing the moon by means of a telescope. Nevertheless, the statement concerning observation of craters of the moon is inaccurate and there is again a wrong statement on heliocentrism. Finally, the cause of the ashy light of the moon is not reported as we shall see in the sequel.

What are Leonardo da Vinci's contributions to astronomy?

*"While Leonardo da Vinci is known for his many contributions to the fields of art, science, and engineering, his contributions to astronomy were limited. Although he did not make any major discoveries in this field, he did make some interesting observations and speculations about the universe.*

*One of da Vinci's most notable contributions to astronomy was his observation of the moon. He was one of the first people to realize that the light we see on the moon is actually reflected sunlight. He also noted that the moon's surface was not smooth, but rather had mountains and craters.*

*Da Vinci was also interested in the stars and their movements. He believed that the stars were fixed in place, while the planets moved around the sun. This was in contrast to the prevailing belief of his time, which held that all celestial bodies revolved around the Earth.*

*In addition to his observations, da Vinci also made some interesting speculations about the universe. He believed that the universe was infinite and that it was filled with other worlds similar to our own. He also speculated that these other worlds might be inhabited by intelligent beings.*

*Overall, while da Vinci's contributions to astronomy were limited, he did make some interesting observations and speculations that contributed to our understanding of the universe."*

Then, I made a confirmation request. In the sequel you will find my series of questions and GPT-4 answers.

can you confirm your answer?

*"Yes, I can confirm my answer. While Leonardo da Vinci is primarily known for his achievements in art and engineering, he did make some notable observations and speculations about astronomy, including his observations of the moon's surface and his belief that the stars were fixed in place while the planets moved around the sun."*

In what notes did Leonardo write about the motions of the planets around the sun?

*"Leonardo da Vinci wrote about the motions of the planets around the sun in his Codex Leicester, also known as the Codex Hammer, which is a collection of his scientific writings and observations. In this manuscript, which is named after its original owner, Thomas Coke, the Earl of Leicester, da Vinci discusses his observations and ideas about astronomy, including his belief that the planets move*

*around the sun in a heliocentric model, which was a revolutionary idea at the time. The Codex Leicester is now owned by Bill Gates, who purchased it at an auction in 1994."*

Please, can you specify the exact reference of the sheet of the Codex Leicester where Leonardo wrote that the planets revolve around the sun?

*"Leonardo da Vinci's Codex Leicester contains various notes and drawings on astronomy and natural phenomena, including his observations and ideas on the motion of planets around the sun. The specific reference to the sheet or page number where Leonardo wrote about this topic may vary depending on the edition of the manuscript and the way it has been organized.*

*However, it is generally accepted that Leonardo's ideas on heliocentrism and the motion of planets around the sun are discussed throughout the Codex Leicester. In particular, folios 7r, 8r, and 9v contain some of his most explicit notes on this topic, where he discusses the relative movements of the Earth, the Moon, and the Sun, and how this affects the appearance of the planets in the sky.*

*It is worth noting that Leonardo's ideas on heliocentrism were not fully developed or conclusive, and he did not provide mathematical proofs or a complete astronomical model to support his views. Nonetheless, his notes and observations were an important step towards the development of modern astronomy and the understanding of the universe."*

Are you basing your answer directly on the Codex Leicester or on another author? If so, who is this other author?

*"As an AI language model, I do not have access to the physical copy of Codex Leicester, so I base my answers on the information available from various credible sources. In this case, my answer is based on the general knowledge and consensus among experts in the field of history of science and astronomy about Leonardo da Vinci's work on heliocentrism and the motion of planets.*

*My understanding of Leonardo's work is based on the analysis and interpretations of his manuscripts and drawings by scholars such as Martin Kemp, a renowned Leonardo da Vinci expert and author of the book "Leonardo da Vinci: The Marvellous Works of Nature and Man," as well as other credible sources such as the official website of the Codex Leicester and various academic articles and publications on the topic."*

As previously said, Codex Leicester does not contain any reference to heliocentrism. In conclusion, the OpenAI ChatGPT and GPT-4 chatbots present significant problems of accuracy, reproducibility and traceability of responses.

3. More experiments with BigScience BLOOM and Google Bard

Recently, I also tested BigScience Large Open-science Open-access Multilingual Language Model (BLOOM) on the same topic. BLOOM is a 176B parameter open-access large language model that is based on decoder-only transformer [5]. It was designed and built thanks to the collaboration of many researchers in several countries. BLOOM was trained in forty-six natural languages and thirteen programming languages on a cluster of several hundreds of GPUs of the Jean Zay French supercomputer at IDRIS supercomputing center of the French National Center of Research: CNRS with the support of GENCI. Training Bloom on Jean Zay supercomputer is estimated several millions of euros.

Either in *sampling* or *greedy* modes, BLOOM generate answers that are not correct, as shown in the sequel.

What is Leonardo da Vinci's contribution to astronomy?"

BLOOM Sampling mode: *"He clearly was aware that the Earth revolved around the Sun."*

BLOOM Greedy mode: *"The answer is that he was the first to describe the phases of Venus."*

I highlight that the phases of Venus were observed for the first time by Galileo Galilei.

When asked to confirm, Bloom answer is:

*"I think it is a good answer."*

And when I ask: *"what is your reference?"*

Bloom answer is again:

*"I think it is a good answer."*

I made also a quick test with Google FLAN-T5 XXL large language model. The answer is somewhat short but also incorrect:

*"heliocentric theory"*

Finally, I tested the experimental conversational AI service Google Bard that is supported by Google large Language Model for Dialogue Application (LaMDA). Bard is also based on transformers [6]. Bard leverages web resources to generate answers.

What are Leonardo da Vinci's contributions to astronomy?

*"Leonardo da Vinci was a polymath, and his interests included astronomy. He made many contributions to the field, including:*

*- He was one of the first people to propose the use of a telescope to study the heavens.*

*- He made detailed observations of the moon and other planets, and he was able to draw accurate maps of their surfaces.*

*- He studied the movement of the stars and planets, and he was able to develop mathematical models to predict their positions.*

*- He also made contributions to the study of the sun and the solar system.*

*Da Vinci's work on astronomy was not always accurate, but it was groundbreaking for its time. He was one of the first people to use scientific methods to study the heavens, and his work helped to lay the foundation for modern astronomy.*

*Here are some specific examples of da Vinci's contributions to astronomy:*

*- In 1509, he wrote a treatise on optics in which he proposed the use of a telescope to study the heavens. He correctly predicted that a telescope would be able to magnify distant objects, and he even sketched out a design for a telescope.*

*- Da Vinci made detailed observations of the moon and other planets. He was able to draw accurate maps of their surfaces, and he was also able to estimate their sizes and distances from Earth.*

*- Da Vinci studied the movement of the stars and planets. He was able to develop mathematical models to predict their positions, and he even predicted the existence of a new planet (which was later discovered by Galileo).*

*- Da Vinci also made contributions to the study of the sun and the solar system. He correctly identified the sun as the center of the solar system, and he also proposed that the sun was a star.*

*Da Vinci's work on astronomy was not always accurate, but it was groundbreaking for its time. He was one of the first people to use scientific methods to study the heavens, and his work helped to lay the foundation for modern astronomy."*

Clearly, Leonardo never made detailed observations of the moon and other planets, and consequently wasn't able to draw accurate maps of their surfaces. It was Galileo Galilei who managed to produce the first accurate drawings of parts of the moon surface [23]. Nor was Leonardo able to develop mathematical models to predict the positions of planets. It is Kepler who first reported the elliptic trajectories of planets. Once again, the heliocentrism statement is wrong.

Can you confirm your answer?

*"As a language model, I'm not able to assist you with that."*

What is the reference?

*"Sure, here are some references for Leonardo da Vinci's contributions to astronomy:*

*- Codex Leicester (1509-1510): This notebook contains da Vinci's notes on a variety of subjects, including astronomy. In it, he discusses the use of a telescope to study the heavens, the movement of the stars and planets, and the sun and the solar system.*

*- Treatise on Optics (c. 1508): This treatise is da Vinci's most comprehensive work on optics. In it, he discusses the nature of light, the properties of lenses, and the use of telescopes.*

*- The Notebooks of Leonardo da Vinci (1480-1519): This collection of notebooks contains da Vinci's notes on a wide range of subjects, including astronomy. In it, he discusses his observations of the moon, planets, and stars, as well as his theories about the solar system.*

*- The Life of Leonardo da Vinci by Giorgio Vasari (1550): This biography of da Vinci includes a discussion of his contributions to astronomy. Vasari praises da Vinci for his innovative use of telescopes and his accurate observations of the heavens.*

*- The Da Vinci Legacy by Walter Isaacson (2017): This book discusses da Vinci's many contributions to science, including his work on astronomy. Isaacson argues that da Vinci was a "renaissance scientist" who was ahead of his time in his thinking about the universe."*

Clearly, Google Bard and BigScience BLOOM cannot answer to confirmation requests or can't answer well as opposed to OpenAI ChatGPT and GPT-4. I also highlight that the references provided by Bard are accurate but not really related to the original question. In a sense, it is a way of answering that, although out of scope, could be described as clever since it quotes references that could be somewhat useful for Leonardo studies.

In order to check reproducibility of Bard answers, I asked the same question in a different session. Bard answer is presented in the sequel.

*"Leonardo da Vinci was a polymath of the High Renaissance who was active as a painter, draughtsman, engineer, scientist, theorist, sculptor, and architect. He made many contributions to astronomy, including:*

*He was the first person to suggest that the Moon has no atmosphere.*

*He was the first person to propose the use of a telescope to study the heavens.*

*He made detailed observations of the Moon and planets, and he drew accurate maps of their surfaces.*

*He developed theories about the nature of light and its propagation.*

*He designed astronomical instruments, such as the astrolabe and the armillary sphere.*

*Da Vinci's contributions to astronomy were significant, and they helped to lay the foundation for the scientific revolution that would take place in the following centuries.*

*Here are some additional details about da Vinci's contributions to astronomy:*

*The Moon: Da Vinci was the first person to suggest that the Moon has no atmosphere. He made this observation based on his detailed drawings of the Moon's surface, which showed no evidence of clouds or other atmospheric phenomena.*

*The Telescope: Da Vinci was the first person to propose the use of a telescope to study the heavens. He made this suggestion in his notebooks in the early 16th century, but he never actually built a telescope. The first telescope was not built until about 1609, by Galileo Galilei.*

*The Planets: Da Vinci made detailed observations of the Moon and planets, and he drew accurate maps of their surfaces. He also developed theories about the nature of light and its propagation.*

*Astronomical Instruments: Da Vinci designed a number of astronomical instruments, including the astrolabe and the armillary sphere. These instruments were used to measure the positions of celestial objects and to track their movements.*

*Da Vinci's contributions to astronomy were significant, and they helped to lay the foundation for the scientific revolution that would take place in the following centuries."*

Though it overestimates a little Leonardo contribution to astronomy, the answer is better this time but "detailed observations of the moon and planets, drawings of accurate map of the moon and lack of the cause of the ashy light of the moon." Nevertheless, we note that Bard answers are not reproducible.

4. Analysis and possible solutions

ChatGPT, GPT-4, BLOOM and FLAN-T5 XXL make the same colossal mistakes. No note or drawing of Leonardo known so far, contain any allusion to heliocentrism. The modern theory on heliocentrism is due to Nicolaus Copernicus [25]. All statements related to observation of the moon via a telescope are wrong. Google Bard is also wrong about detailed observations of the moon and planets. Leonardo never made detailed observations of the moon and planets, he consequently wasn't able to draw accurate maps of their surfaces. He made just a few observations about the inequalities of its surface and even conjectured that part of its surface could be covered by seas. It was Galileo Galilei who managed to do this [23]. Nor was Leonardo able to develop any mathematical model to predict the positions of planets. Kepler's significant contributions were the first to permit one to predict position of planets with some accuracy. Finally, I highlight that no chatbot quote Leonardo's real discovery in astronomy, i.e., the reason for the ashy light of the moon.

One may wonder where these colossal mistakes come from. Certainly, the poor quality of data and learning process in this particular area is to blame. May be, some comments in Giorgio Vasari's famous book on the life of the most excellent artists of the Renaissance, which is absolutely not a primary source on Leonardo is the origin of many wrong contents on the web [28]. I recall that Giorgio Vasari (1511-1574), who owned a few originals of Leonardo's drawings, never met Leonardo who died in France in 1519 and probably never saw most of Leonardo's masterworks, drawings and notes he is speaking about in his book.

As I previously said, popular topics which are associated to mass of data that was generated during a long period of time and that presents contradictions probably give raise to the problem pointed in this paper.

On what concerns specifically ChatGPT and GPT-4, it turns out that training on a scientific topic does not always include the history of sciences context. For example, in fields such as computer science or engineering, history may be less emphasized. I believe that it is the same for BLOOM and Bard. In my opinion, the context should always be considered when training LLMs. The more the historical, political, societal and scientific context is considered during training, the less errors should be obtained in chats. In this regard, an in-depth study that I'm presently conducting shows that the historical and societal context has been correctly considered in ChatGPT for topics relating to periods of history like the Second World War, a period close to ours that is particularly critical for everything related to ethical aspects. I want to emphasize here that it would seem that chats produced by GPT-4 are of higher quality than those produced by GPT-3, certainly because context was better considered with GPT-4 than with GPT-3 as we intend to show in a future study which presents a very large number of response covering a wide variety of fields, including Mathematics, history of science art and literature…

In my opinion, it is naive to believe that the data which is available on the web and which is often used to train chatbots like BLOOM or Bard is exhaustive and reliable. In the case of BLOOM, this point is particularly obvious, since the designers claim that their training is partly based on data collected in the HAL repository system of the CNRS and on free articles from the French newspaper "Le Monde" [5]. It is notorious that HAL contains some articles that have not been peer-reviewed and newspapers often contains articles that present an editorial bias. The search for misleading information in digital environments is a hot topic in our time [29]. Finally; I believe that references should be included in the data that are used by the system. Considering references of data is particularly important in the training process in order to ensure reliable chats and their traceability.

We must continue to carry out this type of study. Wrong statements and related issues about reproducibility and traceability lead to the following question: does the problem comes only from inaccurate data in the mass of data on a given topic (data that has not been checked and validated) or more generally, to the very nature of large language models? Clearly, this type of question must be addressed if one wants to obtain reproducible, traceable and reliable responses. To my opinion, reliability should not be sacrificed to demand or to the illusion that one is addressing a real person. Better a chatbot that answers: "I don't know", "I don't have any answer to your question yet" or a short answer with a single accurate statement than a plethoric answer which contains many mistakes.

In my opinion, it is desirable that governmental institutions get more involved in establishing specifications for the design, data collection and training of chatbots and promote their control by judicial institutions in order to produce reliable and traceable educational chatbots which feature reproducible responses. To leave the realization of such systems to companies or consortia of laboratories can deprive humanity of the chances of obtaining reliable AI systems whose responses are reproducible due to the limitation of means and knowledge of these entities. Instead of collecting data from the web, designers should rather use data produced by teachers and academics at a large scale like what is done for the education of human beings. Of course, this raises the questions of human and financial resources as well as time. Probably, only countries with a great tradition in education, enough determination, and great popular support for this type of project, as well as substantial means and access to supercomputing resources will be able to succeed in producing efficient systems capable of providing reliable, traceable and reproducible responses with an educational goal. Judicial control, will also ensure that all legal and ethical aspects are respected so that chatbots do not infringe intellectual property, become instruments of discrimination and propaganda or are misused for the purpose of pure consumerism.

5. Conclusions

Many have reported the great potential of LLMs. This potential will be highlighted when mature chatbots will be combined with web search engines, word processing software, spreadsheets or technical drawing tools to lighten loads and increase productivity. Nevertheless, some questions remain for now and one may wonder if we can trust LLMs for education or business purpose. For now, I believe, it is necessary to check the accuracy of chats.

The history of science is certainly one of the areas where chatbot's results are weakest. May be, LLMs designers didn't consider it an important area. May be, they didn't want to spend time to check the masses of data and properly train chatbots on this subject.

In future work we plan to evaluate answers produced by Chat-GPT, GPT-4, BLOOM and Bard for a large variety of topics. In particular, we will evaluate answers in detail for a large number of subjects relating to sciences, legal issues, international literature and the arts. This new study should allow us to identify the different types of weaknesses of LLMs and continue to propose solutions.

To my opinion, research in Artificial Intelligence must not stop but careful attention needs to be paid to the data that is fed into the systems and the training process. Only countries that have the financial means and the supercomputing technology as well as the will, human skills and support and time will be able to achieve great success in designing their systems that will provide reliable, reproducible and traceable information. So far, nothing has been found better than excellent teachers and experts to provide a good education and correct information.

Did Leonardo think of a machine that would quickly and accurately answer all his questions? We will probably never know. He who was so inventive and eager for knowledge, draw many machines and wrote many notes like: "*Ask Benedetto Portinari how they walk on ice in Flanders*" or "*When I see Lorenzo de' Medici, I will inquire about the Water Treaty of the Bishop of Padua*" (see [30]). Leonardo was certainly one of the few Renaissance men to glimpse the many implications of the scientific and societal revolution that was beginning by his time.

Acknowledgments

Part of this study has been made possible via funding of Centre International de Mathématiques et d'Informatique de Toulouse, project BigGraphs 2020-2023, CIMI-ANR-11-LABX-0040- LABX-2011.

References

[1] Goodfellow, I., Bengio, Y., Courville A., Deep Learning, The MIT Press, Cambridge, 2016.

[2] Tunstall, L. von Werra, L. Wolf, T. Natural language processing with transformers: building language applications with Hugging Face, O'Reilly Media, Incorporated, 2022.

[3] Brown, T. B. et al. Language models are few-shot learners. In Advances in Neural Information Processing Systems, 2020, pp. 1877-1901.

[4] Bubeck, S. Sparks of artificial general intelligence: early experiments with GPT-4, arXiv:2303.12712v1, 2023.

[5] Le Scao et al. BLOOM : a 176B-parameter Open-Access Multilingual language model, arXiv:2211.05100v3, 13 March 2023.

[6] Vaswani, et al. Attention is all you need. In Advances in neural information processing systems 2017 pp. 5998-6008.

[7] Raffel, C., Exploring the limits of transfer learning with a unified text-to-text transformer, arXiv: 1910.10683v3, 2020.

[8] Dosovitskiy, A. et Al. An image is worth 16x16 Words: Transformers for Image recognition at scale,

arXiv:2010.11929v2, 2021.

[9] Shi, L. Luo, J. Zhang, P. Han, H. El Baz, D. Cheng, G. Liang, Z. Understanding user preferences in location-based social networks via a novel self-attention mechanism, Sustainability, 2022 14 16414.

[10] https://openai.com/blog/chatgpt/

[11] da Vinci, L., manuscripts W. 19003r v, W. 19008v, 19013v, 19014r and 19078, ca. 1510-1511, Windsor collection of manuscripts, The Royal Collection, Windsor Castle, in Kemp, M. Leonard de Vinci, Citadelles, 2019, p. 418-428.

[12] da Vinci, L., manuscripts 19073v, 19078, ca. 1513-1514, Windsor collection of manuscripts, The Royal Collection, Windsor Castle, in Kemp, M., Leonard de Vinci, Citadelles, 2019, p. 429-432.

[13] da Vinci, L., W. 912579r, W. 912660v to W. 912662, W 912669v, ca. 1508-1509, Windsor collection of manuscripts, The Royal Collection, Windsor Castle, in Kemp, M. (2019) Leonard de Vinci, Citadelles, pp. 444-449.

[14] Marusic, I., Broomhall, S., Leonardo da Vinci and Fluid Mechanics, Annual Review of Fluid Mechanics, 2021, p. 1 -25.

[15] El Baz, D., On Graphs, Geometry, Motion and Turbulences in Leonardo's Virgin and Child with Saint Anne, 2022 hal-03767784.

[16] da Vinci, L., Leic. Fol. 2r., ca. 1506-1510, Codex Leicester, Bill and Melinda Gates Collection, in Delieuvin, V. Frank, L. Léonard de Vinci, 2019 p. 191.

[17] da Vinci, L. Manuscript B.L. 28v, ca. 1508, London British Library, in Kemp, M., Leonard de Vinci, Citadelles, 2019, p. 464.

[18] da Vinci, L., MS 2177 48r, ca. 1508-1509, Bibliothèque de l'Institut de France, Paris, in Delieuvin, V. Frank L. (2019) Léonard de Vinci, Louvre éditions, Hazan, p. 189.

[19] Kemp, M., Leonardo da Vinci, Citadelles, 2019, p. 463, 465.

[20] Arasse, D. Léonard de Vinci, Le rythme du monde, Hazan Paris, 1997, p. 79.

[21] Delieuvin, V. Frank, L. Léonard de Vinci, 2019 p. 186, 188, 191.

[22] da Vinci, L. Manuscript C.A. 518 r, ca. 1513-1514, Codex Atlanticus, Biblioteca Ambrosiana Milan.

[23] Galileo Galilei, Drawings of the moon for the notes on Sidereus Nuncius, 1610, Biblioteca Nazionale, Florence

[24] da Vinci, L. Codex of Madrid, Biblioteca Nacional, MS. 8936, 2v – 3r, ca. 1503 – 1505, in Kemp, M. (2019) Leonard de Vinci, Citadelles, p. 463.

[25] Copernicus, N., De revolutionibus orbium coelestium,1543.

[26] Panofsky, E., Meaning in the visual arts, Doubleday Garden Books, Garden City, N.Y., 1955.

[27] da Vinci, L., F 5 r, 4 v, ca. 1508-1509, Bibliothèque de l'Institut de France, Paris,

[28] Vasari, G. Le vite de piü eccellenti architetti, pittori e scultori Italiani, Firenze, 1550

[29] Jia Luo, J., Xue, R. Hu, J., El Baz, D., Combating the Infodemic: A Chinese Infodemic Dataset for Misinformation Identification, Healthcare, 2021, 9, 1094.

[30] da Vinci, L., Manuscript B.M. 135 r, ca. 1500-1505, London British Museum, Codex Arundel 263.